\documentclass [12pt,preprint]{aastex}
\usepackage{epsfig}
\begin{document}
\voffset-1cm
\newcommand{\gsim}{\hbox{\rlap{$^>$}$_\sim$}}
\newcommand{\lsim}{\hbox{\rlap{$^<$}$_\sim$}}

\title{Is there anything special about GRB 080319B?}

\author{Shlomo Dado\altaffilmark{1}, Arnon Dar\altaffilmark{1}
and A. De  R\'ujula\altaffilmark{2}}

\altaffiltext{1}{dado@phep3.technion.ac.il, arnon@physics.technion.ac.il,
dar@cern.ch.\\
Physics Department and Space Research Institute, Technion, Haifa 32000,
Israel}
\altaffiltext{2}{alvaro.derujula@cern.ch; Theory Unit, CERN,
1211 Geneva 23, Switzerland \\
Physics Department, Boston University, USA}

\section*{}

{\bf Gamma-ray bursts (GRBs) are short flashes of gamma rays reaching 
Earth from deep space. They are usually accompanied by a longer-lived 
`afterglow' (AG) emission at lower energies. They are the brightest 
events in the visible universe. On March 19th, 2008, the Swift, Integral 
and Konus-Wind satellites detected$^{1,2,3}$ GRB 0803129B with the 
largest energy fluence, so far, of a long GRB. Three robotic ground telescopes, 
detected its extremely intensive optical light emission$^{4,5,6}$ before 
the Swift alert, and saw it brightening to a visual peak magnitude 5.4, 
visible to the naked eye, some 18 s after the start of the burst. Several 
telescopes continued to follow its AG, which, like its prompt 
emission, is very puzzling$^{7}$ in the `standard' interpretation$^8$ of 
GRBs. Here we show that the properties of GRB 080319B and its AG are well 
reproduced by the `cannonball' model$^{9}$ of long GRBs. It was an 
ordinary GRB, produced by a jet of highly relativistic plasmoids (CBs), 
ejected in a core-collapse supernova (SN) and viewed, as some others 
before, particularly close to the CB-emission axis. It still remains to be 
seen whether GRB 080319B was associated with a SN akin to$^{10}$ 
SN1998bw, the SN type ordinarily associated with GRBs, or with a much more 
luminous SN.}

Swift's prompt alert sent to the world's  
telescopes triggered many follow-up observations. Spectral measurements by 
the VLT  and Hobby-Eberly telescopes determined the GRB's 
redshift$^{11,12}$ to be $z\!=\!0.937$. In the standard cosmology, this implies a 
7.5 Gy look-back time, more than half the age of the Universe, and 6 Gpc 
luminosity distance, $\sim\!6000$ times farther than the most-distant 
object a human eye can see, the Triangulum galaxy M33 with magnitude 
5.7 at a distance of nearly 1 Mpc. At $z\!=\!0.973$, the GRB's peak 
photon energy, $E_p\approx 1260$ keV, its total equivalent isotropic 
$\gamma$-ray energy, $E_{iso}\approx 1.32 \times 10^{54}$ erg, and its peak 
luminosity, $L_p\approx 9.67\times 10^{52}\, {\rm erg\, cm^{-2}}$, 
inferred from the Konus-Wind measurements$^3$, are amongst the largest 
measured for GRBs (see Figs.~1a,1b), but are not unique, e.g., GRBs 
990123, 050717, and 061007 were quite similar.

In the CB model$^{9}$ the prompt $\gamma$-ray and X-ray
emission is dominated by inverse Compton scattering (ICS) of
photons of the `glory' (the early SN light 
scattered away from the radial direction by the pre-SN ejecta). 
Electrons in a CB, partaking of its highly relativistic motion,
Compton up-scatter the glory's
photons to $\gamma$-ray energies and collimate them into a
narrow beam along the CBs' direction of motion. This
relativistic boosting and beaming results in the simple relations
\begin{equation}
E_{iso}\propto \delta_0^3\,;~~~~(1+z)^2\, L_{iso}\propto \delta_0^4\,;~
~~~~(1+ z)\, E_p \propto \gamma_0\, \delta_0\,; 
\label{correlation} 
\end{equation} 
where $\gamma_0$ is the bulk-motion Lorentz factor of a CB, 
$\theta$ is  the angle between the line of sight to the
CB and its direction of motion, and $\delta_0$ is the Doppler factor,
$\delta_0 \approx 2\, \gamma_0/(1+\gamma_0^2\, \theta^2)$,
in an excellent approximation for the relevant large-$\gamma_0$ 
and small-$\theta$ values. 
The strong dependence
on $\gamma_0$ and $\delta_0$ results in correlations
among these observables$^{13}$. The  $[(1 + z)\, E_p,E_{iso}]$
and  $[(1 + z)\, E_p,(1+z)^2\, L_p]$ correlations,
for GRBs with known $z$, are shown in Figs.~1a,1b. They
satisfy the CB-model expectation$^{13}$
depicted as the thick lines. Note that GRB 080319 
is in the expected domain. Other established correlations$^{13,14}$ 
are also well satisfied by this GRB.

The ICS spectrum of the scattered glory's photons is an
exponential cut-off power-law with a spectral
index, $\Gamma\approx 1$, cut-off energy $\approx E_p$, and a
power-law tail$^9$, in agreement with the spectral
index, $\Gamma= 1.01\pm0.02 $ in the 15-350 keV range,
reported$^1$ by the Swift BAT team, and with the Band function
fit to the broader 20 keV - 7 MeV energy range,
reported by the Konus-Wind team$^3$.

For the most probable viewing angles of GRBs, $\theta\approx 1/\gamma_0$, resulting in
$\delta_0\approx \gamma_0$. For small viewing angles, $\theta^2 \!\ll\!  1/\gamma_0^2$,
implying $\delta_0\approx 2 \,\gamma_0$, and local values of
$E_p$, $E_{iso}$ and $L_p$ that are, respectively, 2, 8 
and 16 times larger than the mean 
values for long GRBs, in accordance with the
properties$^3$ of GRB 080319B.
From the explicit
relation $E_p\approx 0.23\,\gamma_0\, \delta_0\,\epsilon_\gamma/(1+z)$ eV,
with  $\epsilon_\gamma\!\sim\! 2$ eV the mean energy of 
the `thin-bremssttrahlung' spectrum of the glory's photons$^{14}$, 
we can estimate $\gamma\! \sim\! 1250$ for this GRB,  close
to the mean$^9$.

The CBs decelerate by gathering and scattering the ISM
particles along their path.  The values of $\delta$ and $\gamma$
stay put at $\delta_0$ and $\gamma_0$
until a `break' time $t_b$, reached when the CB has
swept in a mass comparable to its initial rest mass$^{14}$. Beyond 
$t_b$, the CB begins to decelerate rapidly, and in
a constant-density ISM $\gamma\equiv \gamma(t)$ and $\delta\equiv\delta(t)$ approach a
$\sim t^{-1/4}$ decline$^{14}$.

The AG of a GRB is
dominated by synchrotron radiation (SR) from the electrons of the
interstellar medium (ISM) swept into the CBs and Fermi-accelerated by the 
CBs' turbulent magnetic field. 
The energy flux density, $F_\nu\! \propto\!\nu\,dN_\gamma/d\nu $,
of the AG of a single CB,  
which takes over the prompt emission during the fast 
decline phase of the ICS contribution$^{14}$, is  
\begin{equation}
F_\nu \propto n^{(1+\beta)/2}\,R^2\, \gamma^{3\,\beta-1}\, 
     \delta^{3+\beta}\, \nu^{-\beta}\,,
\label{Fnu}
\end{equation}
where $n$ is the ISM density along the CB's trajectory, $R$ is
the CB's radius, which initially increases at $v\!=\!{\cal{O}}(c)$ 
and approaches a coasting value in a
short time, and $\beta\equiv \Gamma-1$ is the SR spectral index, 
a function of frequency and
time.  

An X-ray AG often has a `canonical' shape:  before the break time
$t_b$, it has a slowly-changing
`plateau phase', beyond which it decays like$^{14}$
$F_\nu \propto t^{-\alpha_X}\, \nu^{-\beta_X}$, with a predicted $\alpha_X=\beta_X+1/2$.
But when $t_b$
precedes the start-time of the observations or 
the end of the prompt ICS-dominated phase, the observed AG has this
power-law decline for starters, and there is no observable break$^{14}$.
In all cases, the SR-determined value of $t=t_b$, is correlated$^{14}$ to the
ICS-dictated values of $E_p$, $E_{iso}$ and $L_p$.  The correlation
is satisfied by GRB 080319B, for which only an upper limit, $t_b\!<\!70$ s, 
can be extracted, as for other very energetic GRBs.

In Fig.~1c we show the X-ray light curve of GRB 080319B measured$^{15}$ with 
the Swift XRT and its CB-model description, assuming a constant ISM 
density and approximating the result by a single dominant or average CB.
The best-fit parameters are $\gamma_0\, \theta=0.07$ and $t_b=70$ 
s, but the fit is good for a smaller $t_b$.
 The AG is reasonably well-fit by a 
power-law with $\alpha_X=1.54\pm 0.04 $, except
around $4 \times 10^4$ s, where 
it is poorly sampled.  As expected$^{14}$  for large 
$E_p$, $E_{iso}$ and $L_p$, no AG break is observed. 
The  temporal index, 
$\alpha_X=\Gamma_X-1/2=1.42 \pm 0.07$, predicted from the late-time 
photon spectral index reported$^{16}$ by the Swift XRT team 
($\Gamma_X=1.92 \pm 0.07$), is 
in agreement with the best-fit temporal index.
At $t\!\sim\!4 \times 10^4$s, the data lie
below the fit. If not a statistical fluctuation, this may be due to
a failure of the constant-density approximation, not surprising
at this level of precision. Better optical data in the same time domain
may resolve this question.

Although ICS dominates the prompt $\gamma$-ray and X-ray
emission$^{9}$, the prompt optical emission 
is dominated by SR, because its flux density
increases with decreasing frequency, $F_\nu\propto
\nu^{-0.5}$, to be compared to the flat ($\beta\approx 0$)
energy flux density produced by ICS.

A GRB starts with a succession of prompt pulses.
Each pulse corresponds to a CB emitted at a time dictated by the chaotic
accretion process that generates them. All properties of prompt individual pulses 
at $\gamma$ and X-ray energies are well described$^9$ 
by ICS of the glory's light, whose photon-number 
density decreases like the density of the `circum-burst' matter, $\propto\!1/r^2$.
The observations of GRB 080319B started too late to 
see the X-ray pulses. But pulses were seen at $\gamma$-ray and optical 
frequencies. Even though the optical pulses are SR-generated, their time
dependence is akin to that of a $\gamma$-ray
pulse, dictated in their rise by the exponentially
increasing transparency of the CB
and the medium, and on their fall by the medium's decreasing density. 
Thus, we fit the optical pulses at a fixed $\bar\nu$ by:
\begin{equation}
F_{\bar \nu}(t) \propto  [e^{-(T/(t-t_0))^{1+\beta}}]\, 
[1-e^{-(T/(t-t_0))^{1+\beta}}],
\label{flare}
\end{equation}
where $T$ is an adjustable time-width and $t_0$ is the start-time
of the pulse. The decline
$F_\nu \propto (t-t_0)^{-(1+\beta)}$ is that implied by Eq.~(\ref{Fnu})
for $n\!\propto\! 1/r^2$,
and the rise, to which the fits are quite insensitive, is described
by the same parameters.

The
optical light-curve of Fig.~1d was obtained by
fitting each of the three early pulses observed by the
TORTORA instrument$^4$ with Eq.~(\ref{flare}). The later-time
AG, given by Eq.(\ref{Fnu}) beyond $t_b\sim 70$ s,
is essentially a power-law decline, insensitive to
 $\gamma_0\, \theta$ and $t_b$ but 
sensitive to $\beta_{opt}$. In the CB model, 
the index $\beta$ is $\sim\!0.5$ below and $\sim\!1.1$
above a decreasing `bend frequency', $\nu_b(t)\!\propto\! n \,\gamma^3\, \delta$, 
which usually `crosses' the optical band within $t\!\sim\! 1$ day, so that
$\beta_{opt}\!\approx\! \beta_X$ thereafter.
Our best fit to the optical AG results in
$\alpha_{opt}=1.40\pm 0.04$, which 
implies a late-time  $\beta_{opt}\approx 0.90$, consistent 
with the late time expectation.
So far no late-time spectral information is available 
to verify this prediction.

When a CB  crosses a density enhancement, 
$\nu_b$ increases due the sudden increase in $n$ 
and the delay in the consequent CB deceleration. The
bend frequency may cross the optical band `backwards': from 
above, to below it. Such a spectral evolution
may have been observed$^7$ some 5000 s
after the onset of the burst. The spectral analysis of the UNLV GRB group$^{17}$
shows a decreased $\beta_X=0.70\pm 0.05$ around that time.
The expected $\beta_{opt}\approx \beta_X-0.5=0.2\pm 0.05$
at that time is consistent with the reported$^7$
spectral evolution around 5000 s after burst.

In Fig.~2d we also show the contribution of a  
SN1998bw-like supernova. If, instead, GRB 080319B was produced by a SN akin to 
SN 2006gy, the most luminous  ever detected$^{20}$, 
its measured R-band peak magnitude 
and its estimated extinction in NGC1260 imply that  
towards the end of July, 2008 it will  reach a J-band peak
energy-flux density of $\sim$4.4 $\mu$Jy.

To conclude, the properties of GRB 080319B, corrected for red-shift effects,
were similar to those of 
other very luminous GRBs, such as 061007$^{18}$ and 050717$^{19}$. We have 
shown that, in the CB model, it was simply an ordinary GRB viewed from very near 
axis, as its main properties  --the prompt observables and the spectral and 
temporal evolution of the afterglow-- are well reproduced.
It remains to be seen whether or not GRB 080319B was generated 
by a supernova akin to SN1998bw, or by a much more luminous one, 
such as SN2006gy. A SN similar to SN1998bw, having exploded around March 
19th, 
2008, at $z=0.937$, should reach peak luminosity
around April 15, 2008. Its brightness, 
however, will be comparable to that of the fading optical afterglow of GRB 
080319B around that time, see Fig.~1d. Such an underlying 
SN may be detected by the most powerful ground-based telescopes via
the change in colours of the optical transient induced by the SN, if the 
host galaxy of GRB 080319B is not too bright. The unprecedented possibility
that GRB 080319B was produced 
by a much more luminous SN should not be difficult to verify.

\noindent

\newpage
\section*{Figure Captions}

\noindent
{\bf Fig.~1}\\
{\bf Top left (a):}  The correlation between the `rest frame' peak photon
energy and the isotropic equivalent total $\gamma$-ray energy of
long GRBs
with known redshift. The thick line is the correlation predicted by the CB
model$^{13}$.
GRB 080319B is indicated by a large star.\\
{\bf Top Right (b):}  The `rest frame' peak photon
energy plotted versus the isotropic peak  $\gamma$-ray luminosity of long
GRBs with known redshift.
The thick line is the correlation predicted by the CB model$^{13}$.
GRB 080319B is indicated by a large star.\\
{\bf Bottom left (c):} Comparison between the light-curve
of the X-ray afterglow of GRB 080319B measured$^{15}$
with the Swift X-ray telescope (XRT) and its CB model description,
Eq.~(2). For very luminous GRBs and a constant ISM density, the
X-ray light-curve  is predicted to have a simple power-law decline.
The best fit temporal decline index, $\alpha_X=1.54\pm 0.04$
and the reported$^{16}$ late-time spectral index, $\beta_X=0.92\pm 0.07$,
satisfy, within errors, the asymptotic CB model
prediction, $\alpha_X=\beta_X+1/2$.\\
{\bf Bottom Right (d):} R-band light curve of
GRB 080319B. Data points are from GCNs quoted in Ref.~7. R-band and
V-band
data were combined by subtraction of 0.26 magnitudes from the V-band data
points. The initial flash is fitted by a sum of three SR
pulses with the shape given by Eq.~(3), beginning at $t_0=4.16,~25.74,~37.38
$ s after trigger, and with widths $T=11.02,~8.68,~4.97$ s, respectively.
The AG is taken over by a late temporal decline  given$^{14}$ by
Eq.~(2), with $\alpha=1.40\pm 0.04$.
Also shown is the contribution to the R-band AG
from a SN akin to SN1998bw$^{10}$,
displaced to the GRB's emission site. \\

\newpage
\begin{figure}[]
\centering
\vspace{-1cm}
\vbox{
\hbox{
\epsfig{file=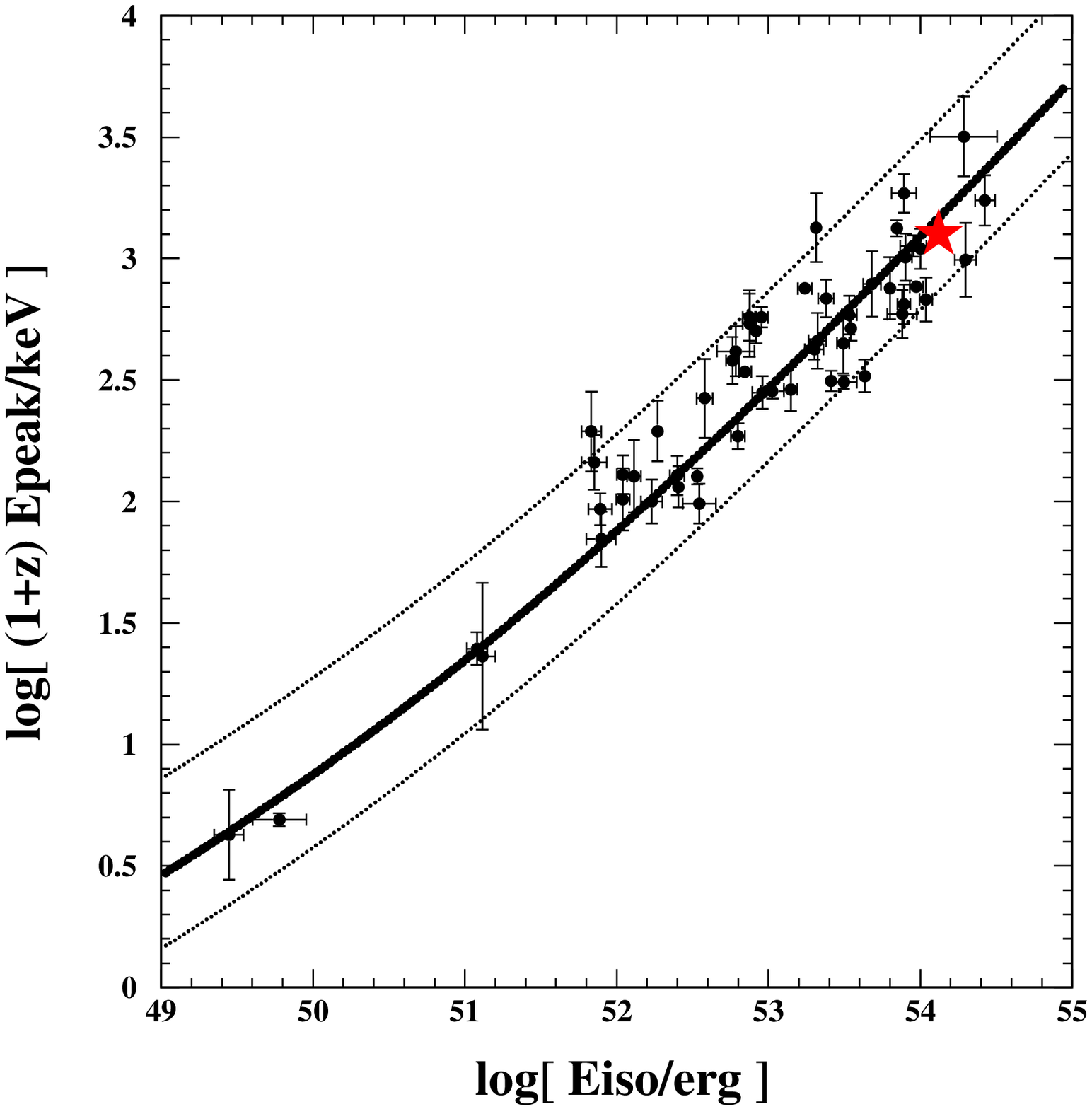,width=8cm}
\epsfig{file=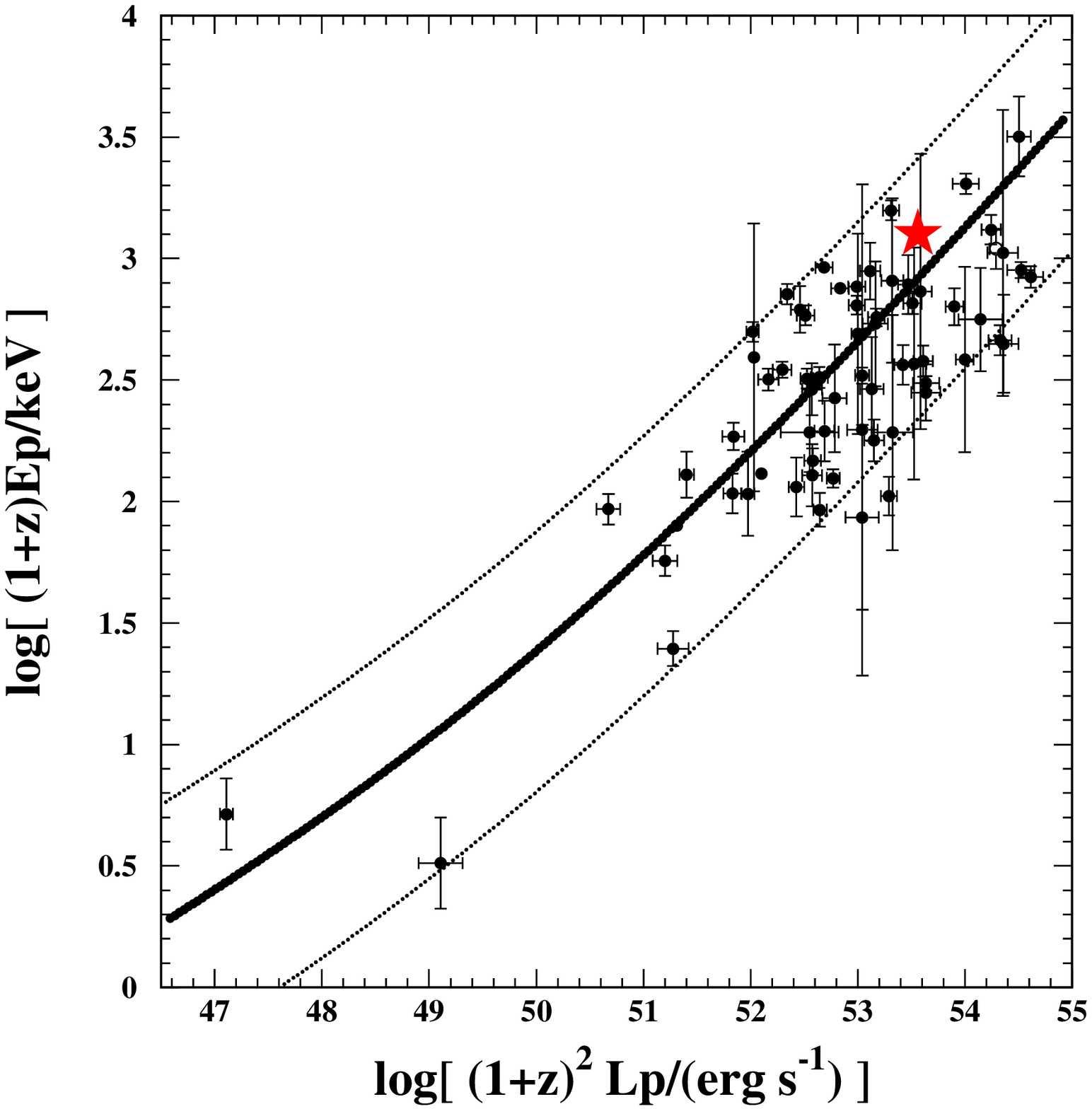,width=8cm}
}}
\vspace{1cm}
\vbox{
\hbox{
\epsfig{file=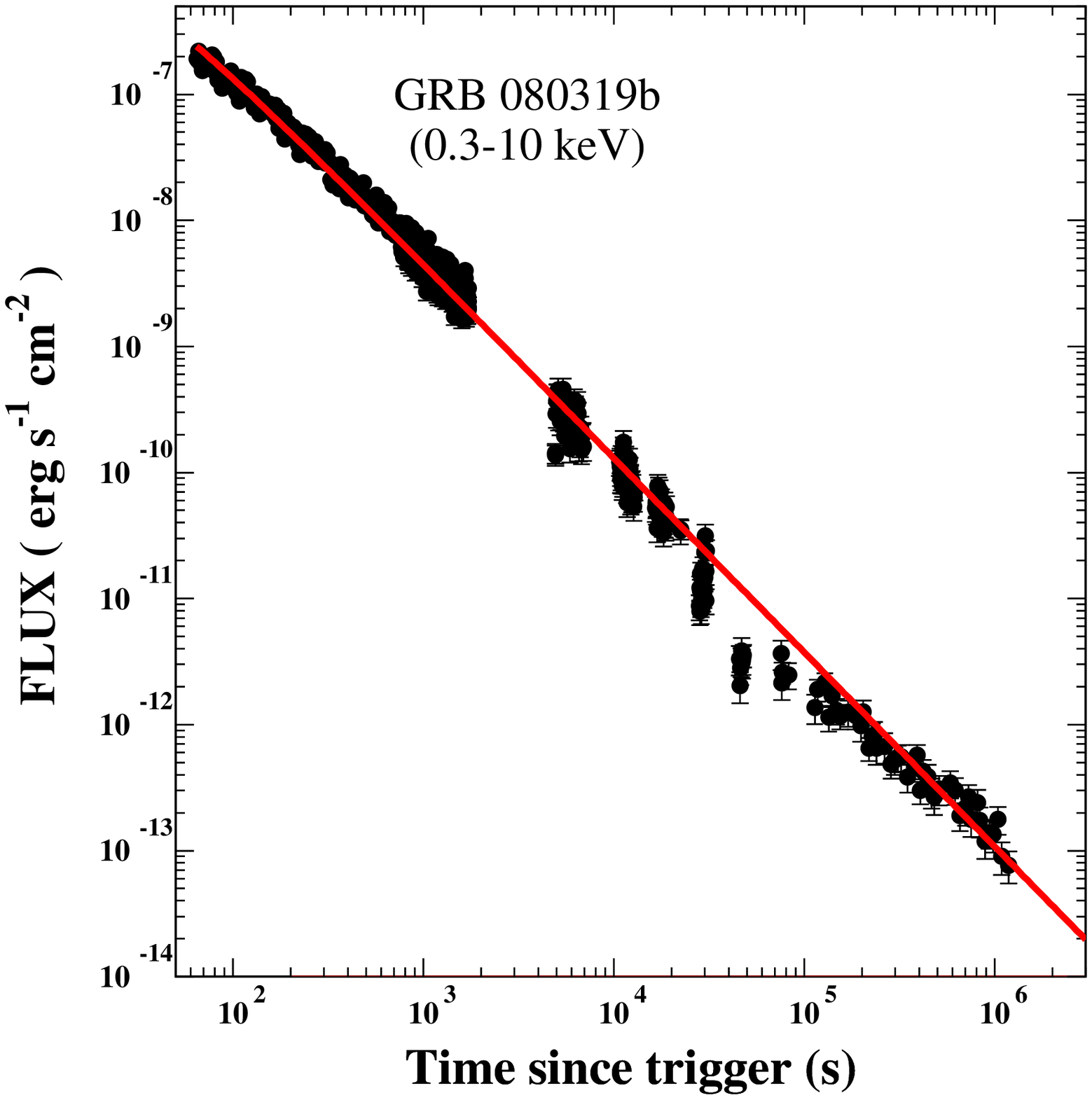,width=8.0cm}
\epsfig{file=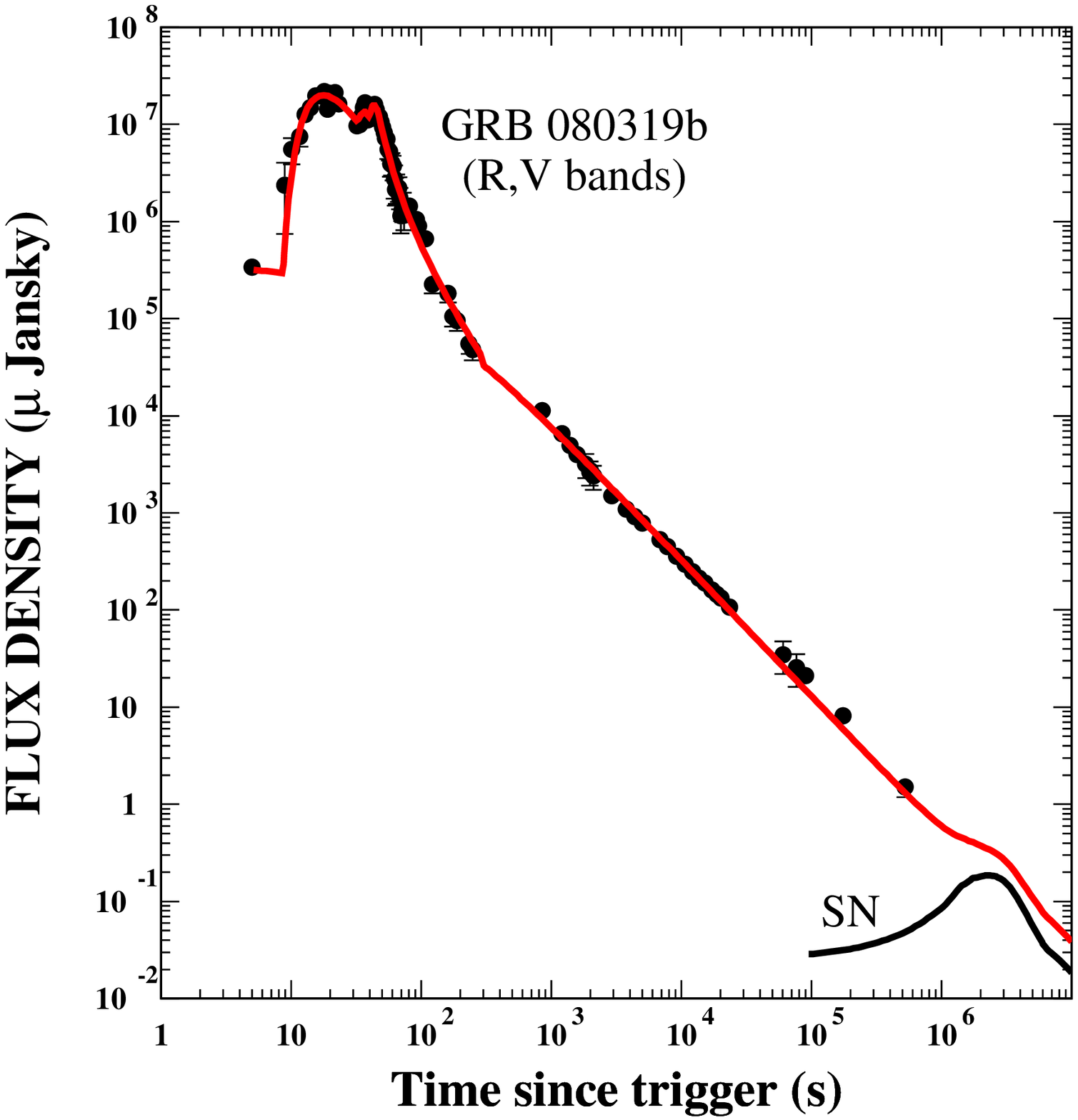,width=8.0cm}
}}
\end{figure}
\end{document}